\pdfoutput=1 
\documentclass{JINST}

\usepackage{caption}   
\usepackage{amsmath}   

\usepackage{enumerate}

\usepackage{subcaption}

\title{The Topo-trigger: a new concept of stereo trigger system for imaging atmospheric Cherenkov telescopes}

\author{Rub\'en~L\'opez-Coto$^{a,b}$,
Daniel~Mazin$^c$,
Riccardo~Paoletti$^d$,
Oscar~Blanch~Bigas$^a$
and Juan~Cortina$^a$\\
\llap{$^a$}Institut de Fisica d'Altes Energies (IFAE), The Barcelona Institute
of Science and Technology, Campus UAB, 08193 Bellaterra (Barcelona)
Spain\\
\llap{$^b$}now at Max-Planck-Institut fur Kernphysik, P.O. Box 103980, D 69029 Heidelberg, Germany\\
\llap{$^c$}Institute for Cosmic Ray Research, University of Tokyo
Kashiwa-no-ha 5-1-5, Kashiwa-shi, 277-8582 Chiba, Japan\\
\llap{$^d$}Universit\`a  di Siena, and INFN Pisa, I-53100 Siena, Italy\\
E-mail: \email{rlopez@mpi-hd.mpg.de}}

\abstract{

Imaging atmospheric Cherenkov telescopes (IACTs) such as the Major Atmospheric Gamma-ray Imaging
Cherenkov (MAGIC) telescopes endeavor to reach the lowest possible energy threshold. In doing so the
trigger system is a key element. Reducing the trigger threshold is hampered by the rapid increase of
accidental triggers generated by ambient light, the so-called Night Sky Background (NSB).
In this paper we present a topological trigger, dubbed Topo-trigger, which rejects events on the basis of
their relative orientation in the telescope cameras. We have simulated and tested the trigger selection
algorithm in the MAGIC telescopes. The algorithm was tested using MonteCarlo simulations and shows a rejection of 85\% of the accidental stereo
triggers while preserving 99 \% of the gamma rays. A full implementation
of this trigger system would achieve an increase in collection area between 10 and 20\% at the energy threshold. The analysis energy threshold of the
instrument is expected to decrease by  $\sim$8 \%. The selection algorithm was tested on real MAGIC data taken with the current trigger
configuration and no $\gamma$-like events were found to be lost.
}

\keywords{Stereo trigger; IACT; MAGIC; Gamma-ray astronomy; Cherenkov telescopes}

\begin{document}

\section{Introduction}

The most violent processes in the universe produce accelerated particles that emit radiation at the highest energies, in the $\gamma$-ray range. They cannot penetrate into the atmosphere due to their interaction with the air molecules. This interaction produces a particle cascade. As the relativistic charged particles produced in the cascade move faster than the speed of light in the atmosphere, they produce Cherenkov light at wavelengths ranging from infrared to UV. The Imaging Atmospheric Cherenkov technique is a powerful procedure to detect the Cherenkov light produced by gamma rays arriving to the atmosphere \cite{iact}.

Lowering the energy threshold of the current generation IACT would allow them to significantly increase the population of different sources (e.g. pulsars, transients, high redshift AGNs). To achieve the lowest possible energy threshold, Imaging Atmospheric Cherenkov Telescopes (IACTs) are built with large reflectors, high quantum efficiency (QE) photomultipliers (PMTs) and fast electronics. The purpose of the study presented in this paper is to decrease the energy threshold of IACTs without increasing the data acquisition rate. We provide results from simulations and piggy-back measurements performed with the MAGIC telescopes. We give a more detailed explanation of the MAGIC telescopes in Section \ref{magic}, where we also present its current trigger system and limitations. We present in Section \ref{topo-trigger} a topological trigger, dubbed Topo-trigger, a new concept of trigger system we consider for MAGIC. Section \ref{telescope_data} describes the piggy-back measurements performed in the telescope and in Section \ref{discussion} we discuss the results and future work to fully implement this trigger system in MAGIC.


\section{The MAGIC telescopes}
\label{magic}

MAGIC is a system of two Imaging Atmospheric Cherenkov Telescopes (IACTs) located in the Roque de los Muchachos, on the island of La Palma (28.8$^\circ$N, 17.9$^\circ$ W at 2225 m a.s.l). The telescopes are dubbed MAGIC~I (M~I) and  MAGIC~II (M~II). The current sensitivity achieved by the telescopes is (0.66 $\pm$ 0.03)\% of the Crab Nebula flux above 220 GeV in 50 hours of data taking at low Zenith distance (Zd) \cite{Performance}. The energy threshold of the telescope is calculated as the peak of the energy distribution of Monte Carlo (MC) $\gamma$-ray events simulated following a Crab Nebula-like spectrum, this is assuming the differential photon spectrum as a power law with the spectral index 2.6. For the events passing all analysis cuts, the energy threshold is $\sim$ 70 GeV. Very-high-energy $\gamma$-ray showers with energies $>$100 GeV produce large amounts of photons easily detectable with MAGIC, but the lower energy ones produce fewer photons, which are harder to detect and to discriminate from triggers generated by NSB. 


\subsection{Trigger system}
\label{trigger_magic}

\begin{figure}[t]
        \centering
  \includegraphics[width=0.45\textwidth]{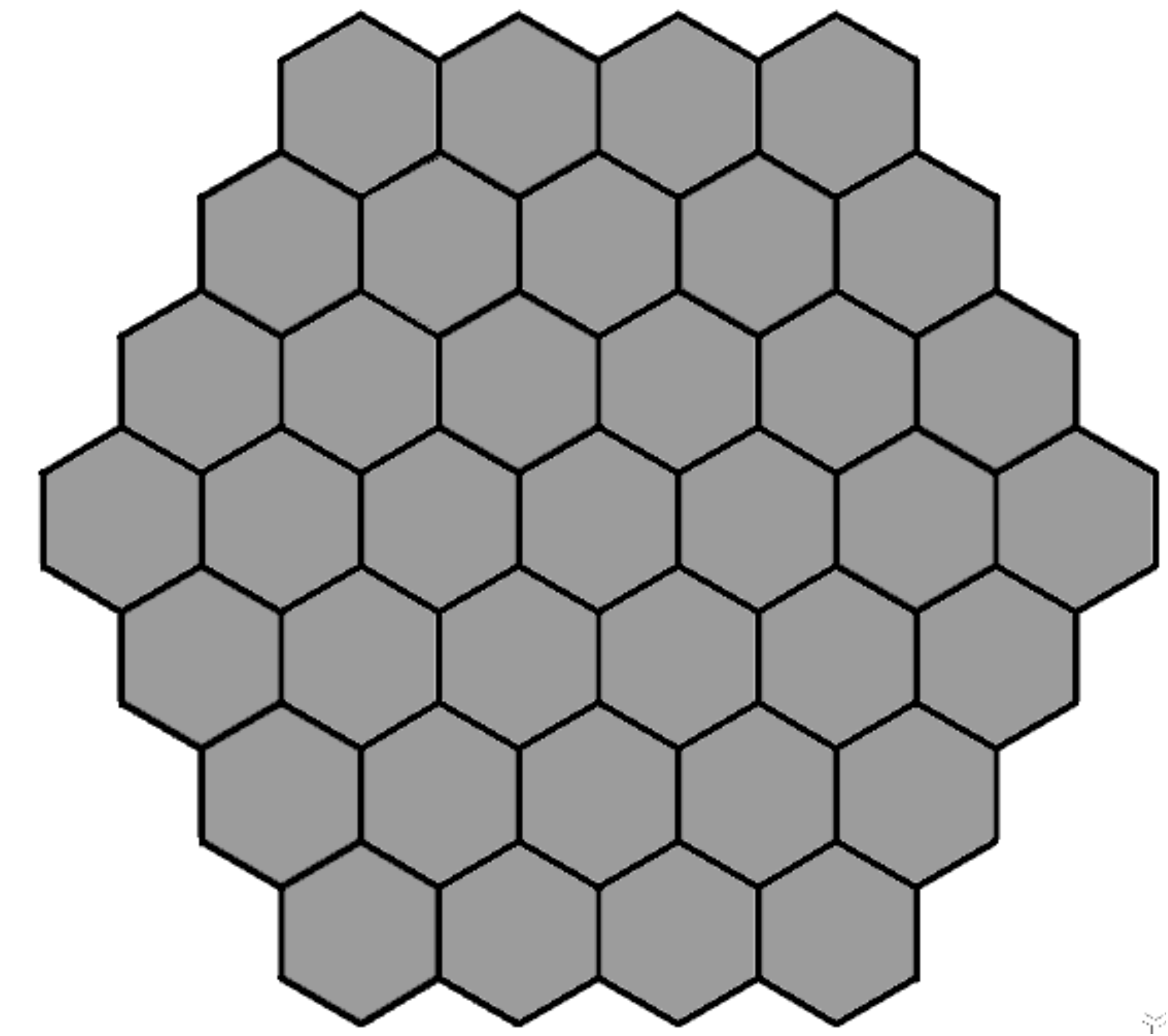}
  \hspace{2cm}
  \includegraphics[width=.4\textwidth]{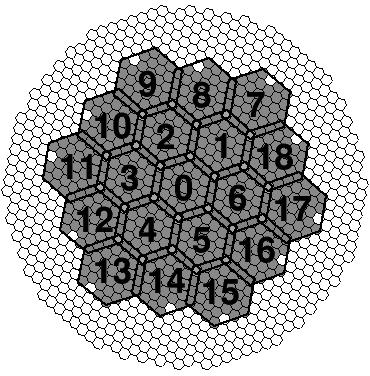}
  \caption{The left panel shows a picture of a single L1 macrocell. The right panel shows is the distribution of the L1 trigger macrocells in the MAGIC camera. The numbers are the ones used in the hardware identification of the macrocells. The lines are imaginary hexagons limiting each of the macrocells.}
  \label{Macrocells}
\end{figure}

The current trigger system installed in the MAGIC telescopes is a digital trigger composed of three stages \cite{trigger_magic}:

\begin{itemize}

\item The \emph{Level 0} (L0) trigger is a pixel trigger that discriminates between the signals that are above or below a determined discriminator threshold (DT) in individual camera pixels. When a pixel is above the threshold, an LVDS signal of 5.5 ns FWHM is sent to the L1 board.

\item The \emph{Level 1} (L1) trigger is a digital filter that searches for spatial and time coincidence of pixels that pass the L0 trigger. For this trigger, pixels are grouped in 19 overlapped cells called \emph{macrocells}. Each macrocell is composed of 36 pixels in an hexagonal configuration as can be seen in the top panel of Figure \ref{Macrocells}. In each of the macrocells, an algorithm searches for n next-neighbour pixels above the threshold. If \emph{n} close-compact next-neighbour (dubbed nNN) pixel signals overlap, the L1 issues a valid trigger from that macrocell. The close-compact configuration for neighbouring pixels is defined when every pixel in the group is in contact with at least other two (except for the 2NN configuration, where by definition every pixel is in contact only with the other one). A valid L1 trigger is issued from every macrocell fulfilling this condition. There are different NN configurations implemented in the L1 trigger (\emph{n} = 2, 3, 4 or 5), although the current one used for stereo observations is \emph{n} = 3, having an effective overlapping trigger gate of 8 -- 9 ns. The output signal of each of the 19 macrocells is processed by a Trigger Processing Unit (TPU) that merges them into an OR gate. 

\item The \emph{Level 3} (L3) receives the output of the TPU and stretches it to a width of 100 ns. It searches then for an overlap between the stretched signals of both telescopes. The effective L3 trigger time window is $\sim$ 180 ns.

\end{itemize}

\subsection{Limitations of the trigger}

The trigger system is hardware limited at several stages. Every time an L1 trigger is issued, the L1 trigger system is busy for 100 ns, not accepting any other trigger during this time. If the L1 rate increases, the time that the system is in a state that can not accept any other trigger, dubbed dead time, starts to be significant. In order not to lose $>2$\% of the cosmic ray events, we cannot accept L1 trigger rates larger than 200 kHz.

\begin{figure}[!h]
  \centering
  \includegraphics[width=0.75\textwidth]{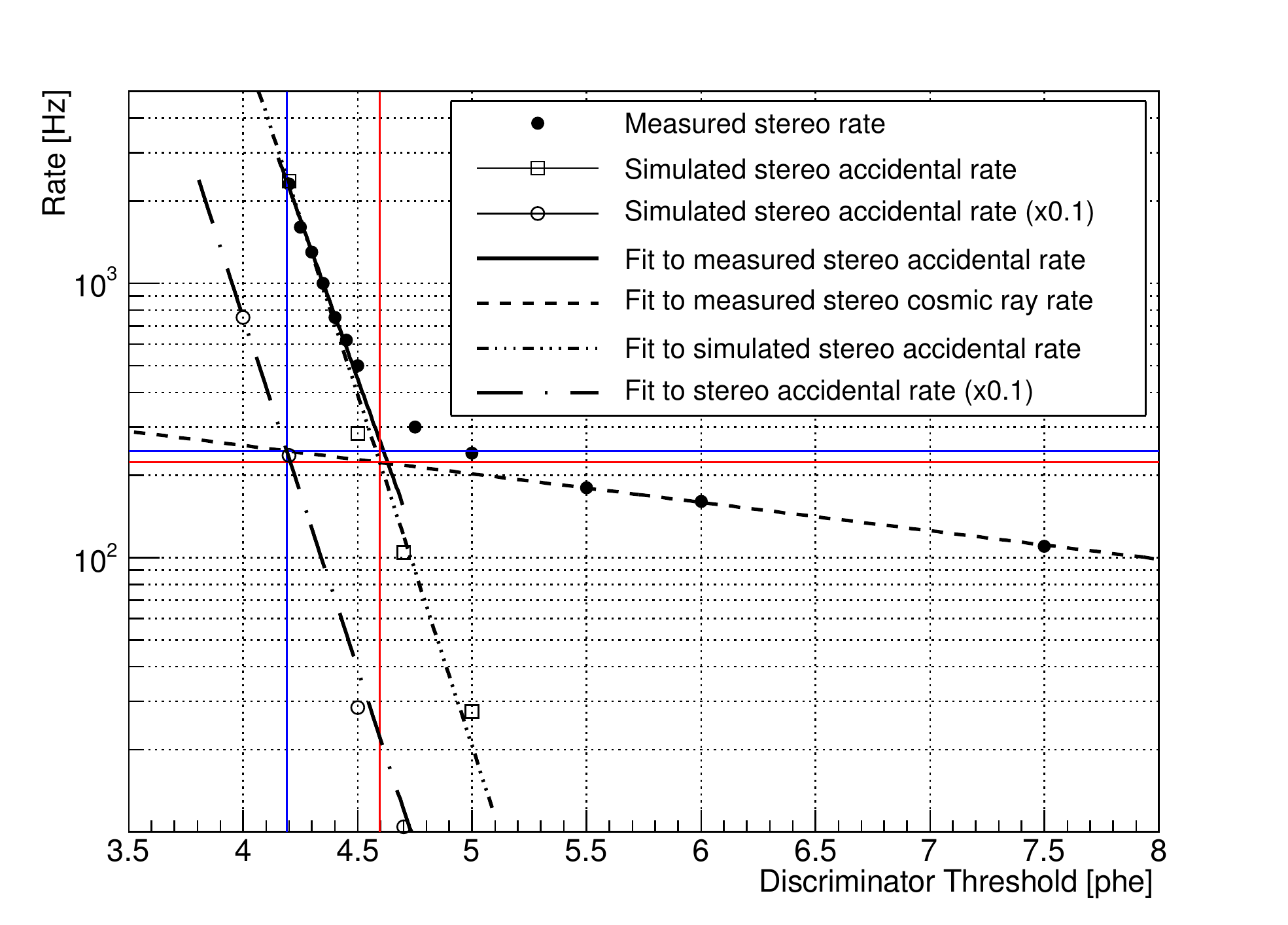}
  \caption{Measured and simulated stereo trigger rate for the MAGIC telescopes. }
  \label{trigger_rate_measured}
  \end{figure}

For the L3, the problem comes from the accidental triggers recorded by the system in coincidence. For a rate dominated by accidental triggers randomly distributed in the camera, the rate of stereo accidental triggers is given by:

$$\textrm{Rate of stereo accidental triggers}=\textrm{L1 rate [M~I]}\times\textrm{L1 rate [M~II]}\times\textrm{L3 Coincidence window}$$

where L1 rate [M~I] is the L1 rate of MAGIC~I, L1 rate [M~II] is the L1 rate of MAGIC~II and L3 Coincidence window is 180 ns. If the L1 trigger rates are 200 kHz, the stereo accidental rate is around 7.2 kHz. This rate can not be recorded by the current Data AcQuisition (DAQ, \cite{DAQ_Diego}) of the telescope, so it has to be decreased. 
The simulated stereo accidental trigger rate (open squares) and measured stereo accidental trigger rate (filled circles) are shown on Figure \ref{trigger_rate_measured}. We are currently working at the crossing point of the extrapolation of the stereo cosmic ray trigger rate and the stereo accidental trigger rate (indicated by the crossing point of the two red lines). With the algorithm we are presenting in this work we aim at reducing the accidental stereo trigger rate the maximum possible. A reasonable goal is to reduce it by a factor 10, marked as open circles in Figure \ref{trigger_rate_measured}. We would then move to operate to the crossing point between the extrapolation of the stereo cosmic ray trigger rate and the 10\% of the accidental stereo trigger rate (the crossing point between the two blue lines).

\section{The Topo-trigger}
\label{topo-trigger}

The trigger logic implemented at this moment in the MAGIC telescope discriminate between showers and NSB using spatial and time information at the single-telescope trigger level and time information at the stereo level. To get rid of additional accidental triggers, we can also use the spatial information at the stereo level.

\subsection{Setup of MC $\gamma$-ray simulations}
\label{simulations}

For the simulation of the Topo-trigger we used the standard MC simulation and analysis software of the MAGIC collaboration \cite{MARS}. In the following we will give a more detailed description. For the atmospheric simulation of gamma rays we used CORSIKA (COsmic Ray SImulations for KAscade) \cite{corsika}. The particles simulated are $\gamma$-ray photons with energies ranging between 10 GeV and 30 TeV, simulated with a power-law function with $\Gamma$=1.6 photon spectral index to have good statistics through the full energy range. For all the calculations, the spectrum was re-weighted to a Crab-like photon spectral index $\Gamma$=2.6. The events are simulated at a $0.4^\circ$ distance from the center of the camera, as it is the standard in MAGIC observations. We also simulated a sample of events at a distance ranging between $0.6^\circ$ and $1.4^\circ$ from the center of the camera to study the performance for off-axis events. The Zd ranges from 5 to 35 degrees because the lowest possible energy threshold is achieved when the telescope is pointing close to the zenith. The Azimuth (Az) angle ranges from 0 to 360 degrees and the maximum impact parameter simulated is 350 meters from the center of the array because there are no showers of this energy range and at those Zd triggered with larger impact parameter. We used 3$\cdot$10$^{6}$ showers. The output of CORSIKA is the information of the direction and position on the ground of the Cherenkov photons produced in the shower. It is then treated with the  \emph{Reflector} program, that calculates the amount of light reflected by the telescope dish and the number of photons arriving at the camera plane. From there on, the \emph{Camera} program simulates the readout system, including PMT response, trigger and data acquisition.

The average Discriminator Threshold (DT) currently applied to the PMTs in the MAGIC simulations (Nominal DT) is 4.5 phe for MAGIC~I (M~I) and 4.7 phe for MAGIC~II (M~II). Different DTs are used for the two telescopes essentially because the mirrors have different reflectivities and the PMTs have different QE. We also simulated reduced DTs to evaluate the potential of the Topo-trigger. We selected DTs that lead to one order of magnitude higher accidental rate after the stereo coincidence of the L3 trigger. When applying the Topo-trigger concept, this accidental rate is reduced, and we evaluated the reduction. We reduced the DT to 4.2 phe in M~I and 4.3 phe in M~II ({Reduced} DT). The results for the stereo accidental trigger rates for { Reduced} and { Nominal} DTs can be found on Table \ref{tab:L1_rates}. 


\begin{table}[t]
\centering
\begin{tabular}{c|c|c|c}

\multicolumn{1}{c|}{} & \multicolumn{3}{|c}{Accidental trigger rate [kHz]}\\ \hline
 DT   & M~I  & M~II  & Stereo\\ \hline\hline

Nominal & 25 $\pm$ 4 & 39 $\pm$ 5  & 0.18 $\pm$ 0.04\\ \hline

Reduced  & 78 $\pm$ 7 & 125 $\pm$9 & 1.8$\pm$ 0.2\\ 

\end{tabular}
\caption{L1 trigger rates for different DTs and for the two MAGIC telescopes. The Nominal DTs correspond to 4.5 phe for M~I and 4.7 phe for M~II, while the Reduced DTs correspond to 4.2 phe for M~I and 4.3 phe for M~II.} 
\label{tab:L1_rates}
\end{table}


\subsection{Spatial information available at trigger level}

\begin{figure}[!h]
 \centering
 \includegraphics[width=0.75\textwidth]{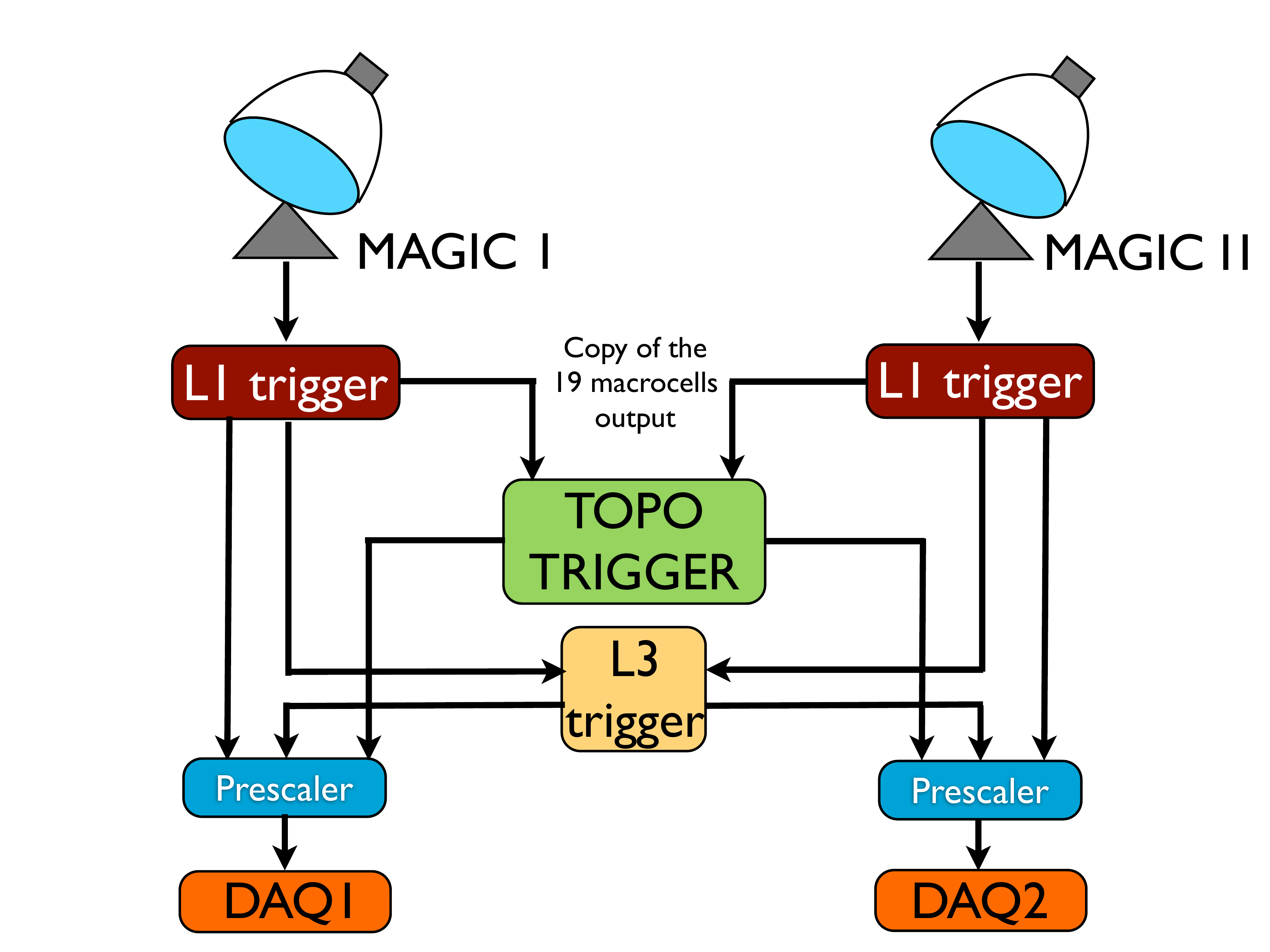}
\caption{Scheme of the MAGIC trigger system including the implementation of the Topo-trigger.}
  \label{trigger_scheme}
 \end{figure}

There have been several attempts to reduce the energy threshold of IACTs, both at the single-telescope and stereo level \cite{topo-veritas,topo-veritas2,level2-magic}. A stereo trigger makes use of the combined information of the triggers in two (or more) telescopes. To apply a cut here has the advantage that the events have already passed through three different levels of trigger, therefore the number of accidental triggers has already been decreased. The basic idea of the Topo-trigger is to implement online cuts on the trigonometric information available at the trigger level. In particular, we want to use all 19 L1 trigger macrocell bits from the two telescopes to define efficient accidental rejection cuts. A scheme of the MAGIC trigger system including the future Topo-trigger is shown on Figure \ref{trigger_scheme}. When an L1 trigger is issued in each telescope, a copy of the signal goes to the prescaler and another one to the L3 trigger. We intend to deliver another copy to the Topo-trigger. The output of the L1, L3 and Topo-trigger go to a prescaler board, that handles all the triggers and tags the events depending on which trigger type they issue. The Topo-trigger compares the macrocells triggered in each telescope every time an L1 trigger is issued and sends a veto signal to the prescaler when the combination of macrocells does not correspond to that triggered by a gamma ray.

 \subsection{Macrocell selection}
 \label{macrocell_selection}
\begin{figure}[t]
  \centering
  \includegraphics[width=0.75\textwidth]{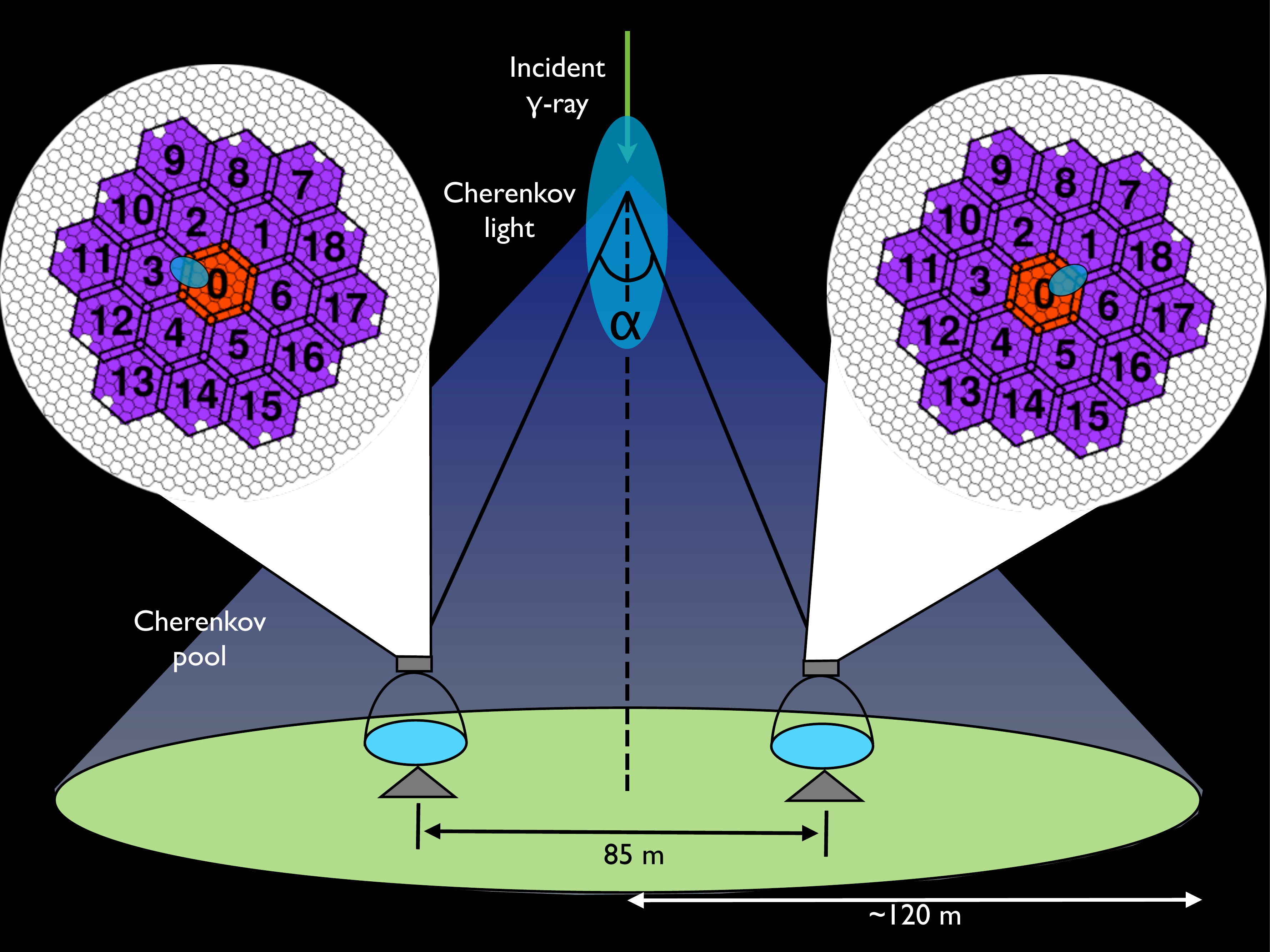}
  \caption{Scheme of the detection of Cherenkov light produced by a low energy $\gamma$-ray shower by the MAGIC telescopes. The angle $\alpha$ between the light arriving to both telescopes is smaller than $0.6^\circ$ for all the showers produced at a height of $\sim$ 10 km.}
  \label{Shower_macrocells}
 \end{figure}

 Let us discuss the separation angle, under which the shower is seen from the two MAGIC telescopes. We denote this angle alpha, see Figure \ref{Shower_macrocells}. The angle is maximum when the shower develops in between the two telescopes. The distance between M~I and M~II is 85 m, and the assumed distance at which the shower is produced is 10 km a.s.l. ($\sim$7.8 km above the telescopes), therefore the maximum angle separation between a point-like shower between the two telescopes will be $\alpha\simeq0.6^\circ$. With the macrocell distribution shown on Figure \ref{Macrocells} and since every pixel covers 0.1$^\circ$, we conclude that the maximum separation of the showers seen in the two MAGIC cameras is 1 macrocell. We have to point out that the angle $\alpha$ calculated depends on the height at which the shower interacts with the atmosphere (high energy showers go deeper into the atmosphere, therefore produce larger $\alpha$ angles). We are also assuming that the same part of the shower is illuminating both MAGIC cameras. If due to fluctuations, different parts of the shower illuminate the cameras, the angle may vary as well.

\begin{figure*} [!htb]   
\begin{minipage}[]{0.5\textwidth}
\includegraphics[width=\linewidth]{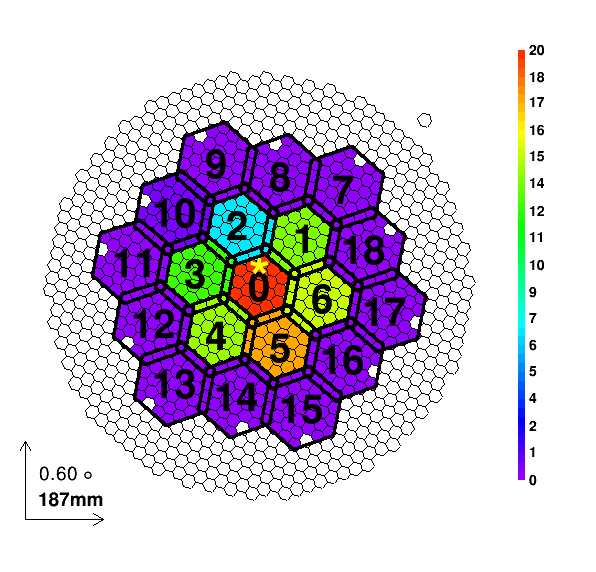}
\end{minipage}
\hspace{\fill}
\begin{minipage}[]{0.5\textwidth}
\includegraphics[width=\linewidth]{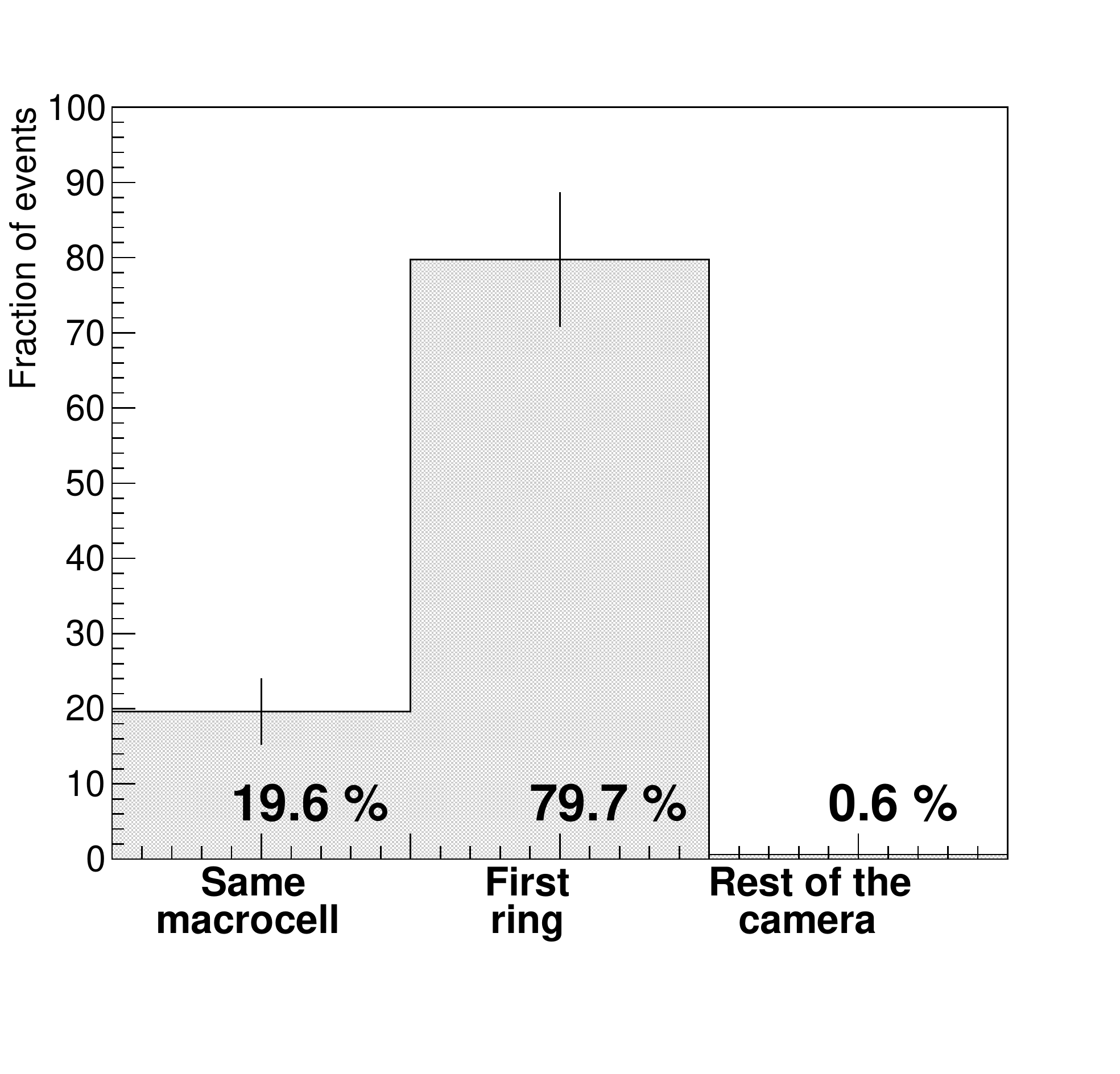}
\end{minipage}

\vspace*{0.5cm} 
\begin{minipage}[]{0.5\textwidth}
\includegraphics[width=\linewidth]{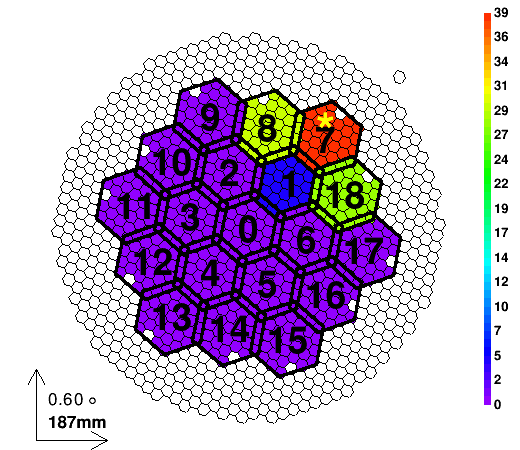}
\end{minipage}
\hspace{\fill}
\begin{minipage}[]{0.5\textwidth}
\includegraphics[width=\linewidth]{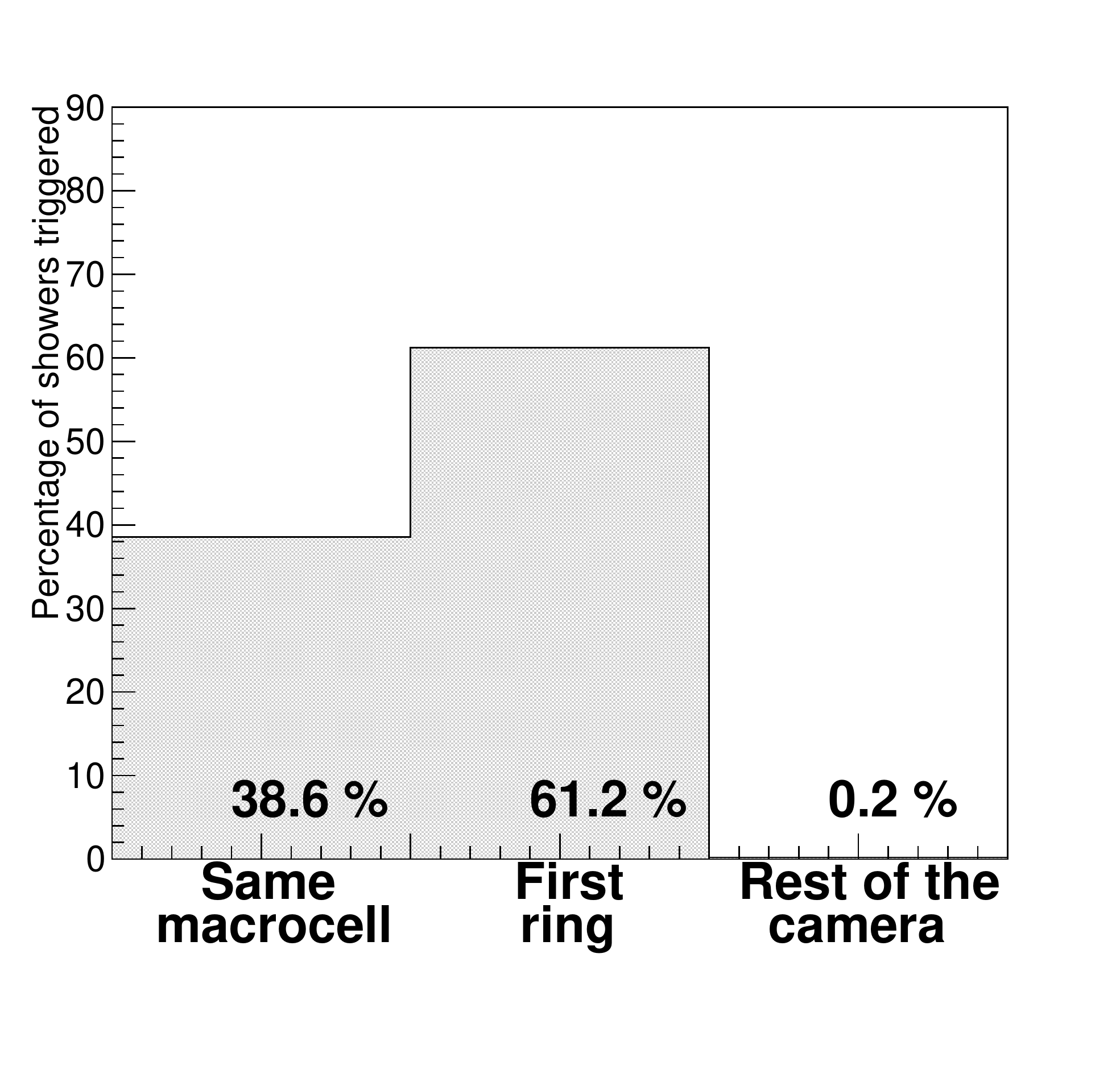}
\end{minipage}
 \caption{Macrocells triggered in M~II when selecting one macrocell in M~I for events triggering only one macrocell in each telescope (left panels) and the histograms of the distributions (right panels). Results were obtained using simulated gamma rays after analysis cuts with Az angle ranging from 0 to 360 degrees. The radial and azimuth asymmetry seen in the macrocell distribution comes from building the stereo trigger with only two telescopes. }
 \label{Macrocells_and_histograms}
\end{figure*}

The selection algorithm implemented in the simulation works as follows: we record the macrocell digital output for 10 ns after the L1 trigger is issued, which accounts for the arrival time difference of the shower to both telescopes. This digital output is 0 if the macrocell was not triggered during those 10 ns and 1 if it was triggered. As the accidental triggers are random, we expect them in general to trigger only one macrocell. Using MC simulations, we actually calculated that  the probability that an accidental event is triggered by more than one macrocell is $\text{P}_{2\text{M}}=0.4\%\cdot\text{P}_{1\text{M}}$, where $\text{P}_{2\text{M}}$ is the probability of triggering 2 macrocells due to an accidental and $\text{P}_{1\text{M}}$ the probability of triggering 1. As the fraction of events triggering more than one macrocell is much smaller than the one triggering only one, we selected the events that triggered only one macrocell in each telescope and studied them. Let us consider the combinations of macrocells in the two telescopes. We select a shower that triggers a given macrocell in M~I and look at the macrocell distribution for these showers in M~II. Figure \ref{Macrocells_and_histograms} shows two examples: in the top panel, macrocell 0 (the central one, marked with an asterisk) is triggered in M~I, and on the bottom panel macrocell 7 (one of the border ones) is triggered. The color scale corresponds to the percentage of events that triggered the macrocell marked with an asterisk in M~I and the corresponding colored macrocell in M~II. In the right plots we show the distribution of the distance between the macrocell triggered in M~I and the macrocell triggered in M~II. The first bin corresponds to the events triggering the same macrocell in M~I and M~II (macrocell 0 or 7). The second one corresponds to events triggering a macrocell in M~II that surrounds the macrocell triggering in M~I (either macrocells 1, 2, 3, 4, 5 and 6, or macrocells 1, 7, 8 and 18. The third bin corresponds to the rest of the macrocells. The y-axis represents the fraction of the total 1--1 combinations triggered on each of the regions. We found that the same happens when we evaluate other macrocells in the first telescope: gamma rays hit either the same macrocell as in the first one or the surrounding macrocells (1 ring) only. The power of the Topo-trigger lies on the fact that the triggers produced by $\gamma$-ray showers always result in certain macrocell combinations, while the accidental triggers produced by NSB produce random combinations. Accepting only the combinations produced by $\gamma$-ray showers allows to reject a large fraction of accidental triggers without losing on $\gamma$-ray efficiency.

Each camera has N=19 macrocells, so there are N$^2$=361 possible combinations of a macrocell in M~I and a macrocell in M~II. Since we only accept events triggering the same macrocell or the surrounding ones in each telescope, we would accept a total of 103 different combinations. Gamma-ray events are distributed according to the distribution we have obtained with the simulations, but the triggers due to accidentals will be randomly distributed in the whole camera, and most of them will be triggering only one macrocell each time. Bright stars could increase the accidental trigger rate on a non-random way, but an automatic individual pixel rate control avoids it by increasing the DTs of pixels affected by stars \cite{Upgrade}. It becomes clear that if we reject the triggers produced in the macrocell combinations that are not fulfilling the condition of being the same macrocell or the surrounding ones in both telescopes, we would reject:

$$\genfrac{}{}{0pt}{}{\mathrm{Fraction\ of\ accidental\ stereo}} {\mathrm{\ triggers\ rejected}}=\frac{361-103}{361}\times100=71\%$$

\subsubsection{Zenith/Azimuth selection}
\label{supertopo}
We studied the Zd and Az dependence 
of the macrocell selection. 
First, we divided data in 4 different Zd bins (equidistant in cos[Zd])
and looked for differences in macrocell distributions
(see Fig. \ref{Macrocells_and_histograms} for an example). It turned out 
that there is no significant difference in distribution of macrocells
as a function of the Zd. 
However we find significant differences in hit fraction of the neighboring macrocells
depending on the Az angle of the observation, since it changes the relative orientation of the two MAGIC telescopes in respect to the showers.
We found that depending on the Az angle, some neighboring macrocells do not contain hits of gamma-ray showers. Those macrocell combinations can be excluded,
which further increases the rejection power of the accidental triggers.
We find that having 12 bins in Az is optimal
since more bins does not increase the rejection power 
(the macrocell combinations do not differ sufficiently by increasing number of bins).

Summarizing we decided to use a single bin in Zd and 12 equidistant Az bins for macrocell
combinations. The tables 
with the macrocells selected in one telescope
depending on the one triggering the other and the
Az angle at which the telescopes are pointing can
be found in \ref{super-topo tables}. The fraction of rejected accidental events is also independent of the DT applied. The average number of macrocell combinations selected after dividing the data in Az bins is 53, therefore the fraction of accidental events rejected is given by:

$$\genfrac{}{}{0pt}{}{\mathrm{\%\ Accidental\ stereo}} {\mathrm{\ triggers\ rejected}}=\frac{361-53}{361}\times100=85\%$$


\subsubsection*{Off-axis simulations}
In the previous sections we have verified that the selection algorithm works applying it to the standard MC simulations with the source located at a 0.4$^\circ$ distance from the center of the camera. Now we want to check that the algorithm does not depend on the source position of the camera, but on the relative position of the telescopes with respect to each other.
We simulated $\gamma$-ray events at distances from the center of the camera ranging from 0.6$^\circ$ to 1.4$^\circ$. We did not include NSB in our simulations to study the effect on events triggered by a $\gamma$-ray shower. We examined the triggered macrocell distribution and applied the Topo-trigger macrocell selection mentioned in Section \ref{macrocell_selection}. The distribution of triggered macrocells is similar to the one obtained using the standard MC with the source at 0.4$^\circ$ from the camera center.

\subsection{Expected performance}
\label{performance}
We will now calculate the collection area and energy threshold of the instrument after applying the Topo-trigger selection.  For the analysis of the MC $\gamma$-rays, we calibrate the data, clean the images applying the image cleaning (IC) method and apply a $\gamma$/hadron separation with the so-called random forest algorithm as in the standard MAGIC analysis \cite{MARS}. We ran MC simulations for two cases, and for the second case we apply two different ICs:

\begin{enumerate}[a)]

\item Nominal DT with standard IC. 

\end{enumerate}

\begin{enumerate}[b.1)]

\item Reduced DT with the standard IC (6 phe for core pixels and 3.5 phe for the neighbour ones).

\item Reduced DT with an IC with charge parameters reduced by 7 \% (which is the mean DT reduction applied from Nominal to Reduced DT). This example is to illustrate the improvement we could achieve if we were able to further reduce the parameters used for the IC.

\end{enumerate}

\begin{figure}[t]
  \centering
  \includegraphics[width=.75\textwidth]{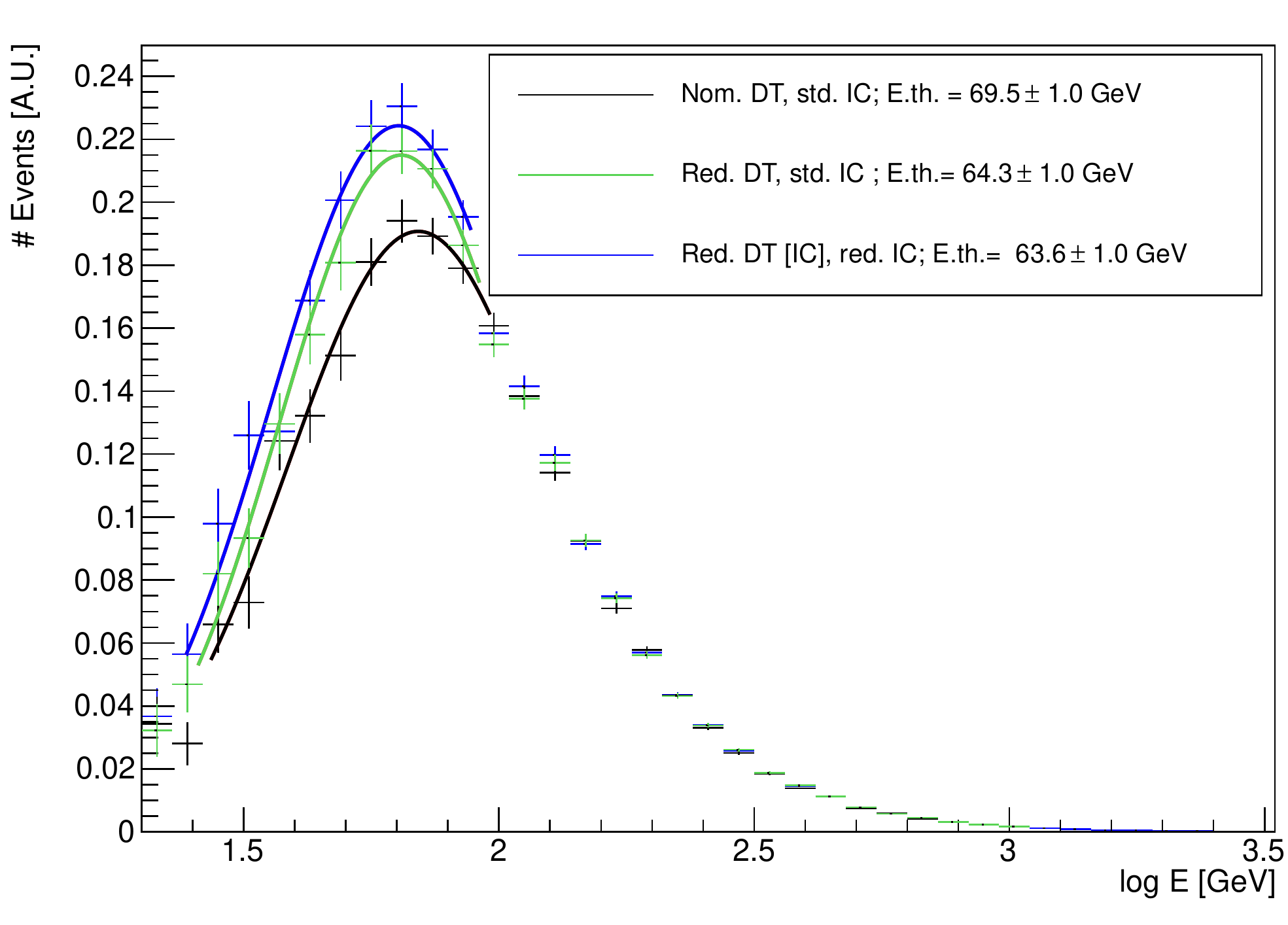}
  \caption{True energy distribution of MC gamma rays (in arbitrary units) for a source with a 2.6 spectral index for different trigger configurations. The lines drawn are fits to gaussian functions for every dataset. }
  \label{Energy_threshold_Turku}
 \end{figure}

Figure \ref{Energy_threshold_Turku} shows the true energy distribution of MC $\gamma$-ray events for the different trigger configurations. The energy threshold is defined as the peak of this distribution and to determine it we fit a gaussian function to the data. The black histogram represents the rate for the current trigger configuration of MAGIC (Nominal DT). The green histogram represents the rate for the Reduced DT configuration with DT=4.2 phe for M~I and DT=4.3 phe for M~II applying Topo-trigger macrocell selection with the standard IC. The blue line represents the rate for the Reduced DT configuration with a 7\% reduced IC. We can see that the energy threshold goes down by up to 8 \% at the analysis level.

\begin{figure}[t]
  \centering
  \includegraphics[width=.75\textwidth]{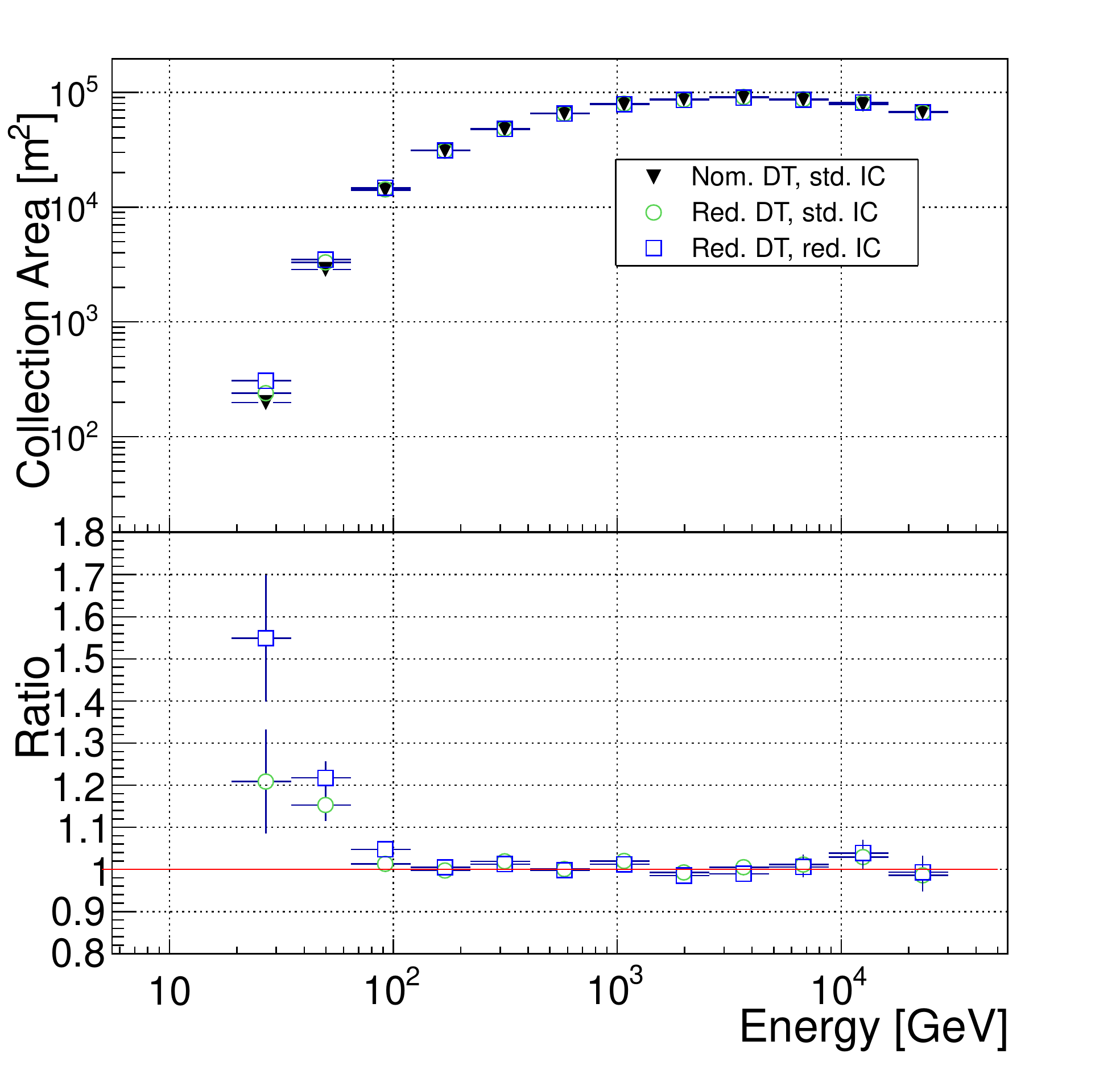}
  \caption{Collection area using the Nominal DT and the Reduced one applying cuts (top) and the ratio between the Reduced DT configurations and the Nominal DT one (bottom). }
  \label{collection_Turku}
 \end{figure}

The collection area in each energy bin is calculated using the following formula:

$$\text{A}_{\text{col}} [\Delta \text{E}]= \frac{\text{N}_{\text{trig}}[\Delta \text{E}]}{\text{N}_{\text{sim}}[\Delta \text{E}]} \text{A}_{{\text{max}}}[\Delta \text{E}]$$

where $\text{A}_{\text{col}}$ is the collection area, $\text{N}_{\text{trig}}$ is the number of triggered events that survive the $\gamma$/hadron separation cuts, $\text{N}_{\text{sim}}$ the number of simulated events and $\text{A}_{\text{max}}$ the maximum simulated area, all of them for the energy bin $[\Delta \text{E}$]. The collection area for both the Nominal DT and the Reduced one are shown on the top panel of Figure \ref{collection_Turku}. The ratio of the collection area obtained using the current MAGIC trigger configuration and the collection area obtained with the Reduced DT applying the Topo-trigger macrocell selection is shown on the bottom panel of the same figure. The black points correspond to the collection area obtained with the current trigger configuration of MAGIC. The green points correspond to the collection area obtained reducing the DT to 4.2 phe in M~I and 4.3 phe in M~II and applying the Topo-trigger macrocell selection. The blue points represent the rate for the Reduced DT configuration with a 7\% reduced IC. The red line in the bottom panel delimits the region where the Reduced DT option performs better than the Nominal DT (ratio $>$ 1). We can see that the improvement in collection area using the Topo-trigger  at the lowest energies is $\sim$60\%, although the errors are very large due to the low statistics at these energies. At the energy threshold, where we have a peak in the number of events triggered, the improvement using the Topo-trigger is between 10--20 \% with respect to the current MAGIC configuration. Moreover, this gain in events is still $\sim$5\% for showers with energies between 70 GeV to 100 GeV. No change, as expected, for higher energies. At the highest energies (the last point in energy of Figure \ref{collection_Turku}), there are only a few events that give trigger, therefore the fluctuations are large. We would like to stress that all these improvements are at the analysis level, i.e. for the events that are used to derive spectra, light curves and skymaps. One also has to keep in mind that these results were obtained using simulations and the results using real data might change.

After applying the Topo-trigger cuts at the trigger level we keep 97.6 \% of the total of triggered $\gamma$-ray events. If we apply the Topo-trigger cuts at the analysis level, we keep 98.8\% of the total events that survived the analysis cuts, meaning that most of the events that were rejected when applying the macrocell algorithm, are not used for analysis either. Lower DTs down to a level of 4 phe were also tested and a percentage of rejected gamma rays at a level of 1--2 \% is constant for any of the DTs tested.

\section{Piggy-back measurements}
\label{telescope_data}

We installed a system to record the macrocell pattern for every event at the MAGIC telescope. The DT of the system could not be lowered because the device installed could not apply any veto to the signals recorded. As the L0 rate was not increased, we do not expect any improvement by applying the algorithm to the data. We cannot make a direct comparison of the data and MC simulations because in the data, apart from $\gamma$-ray events, we also have hadrons, electrons and diffuse gammas. What we seek with these measurements is to validate that the Topo-trigger algorithm does not worsen the performance of the telescope working for the standard operation. To do it, we verify if any $\gamma$-like events are lost at the analysis level due to the Topo-trigger macrocell selection.

\subsection{Description of the setup}

\begin{figure}[t]
 \centering
 \includegraphics[width=0.75\textwidth]{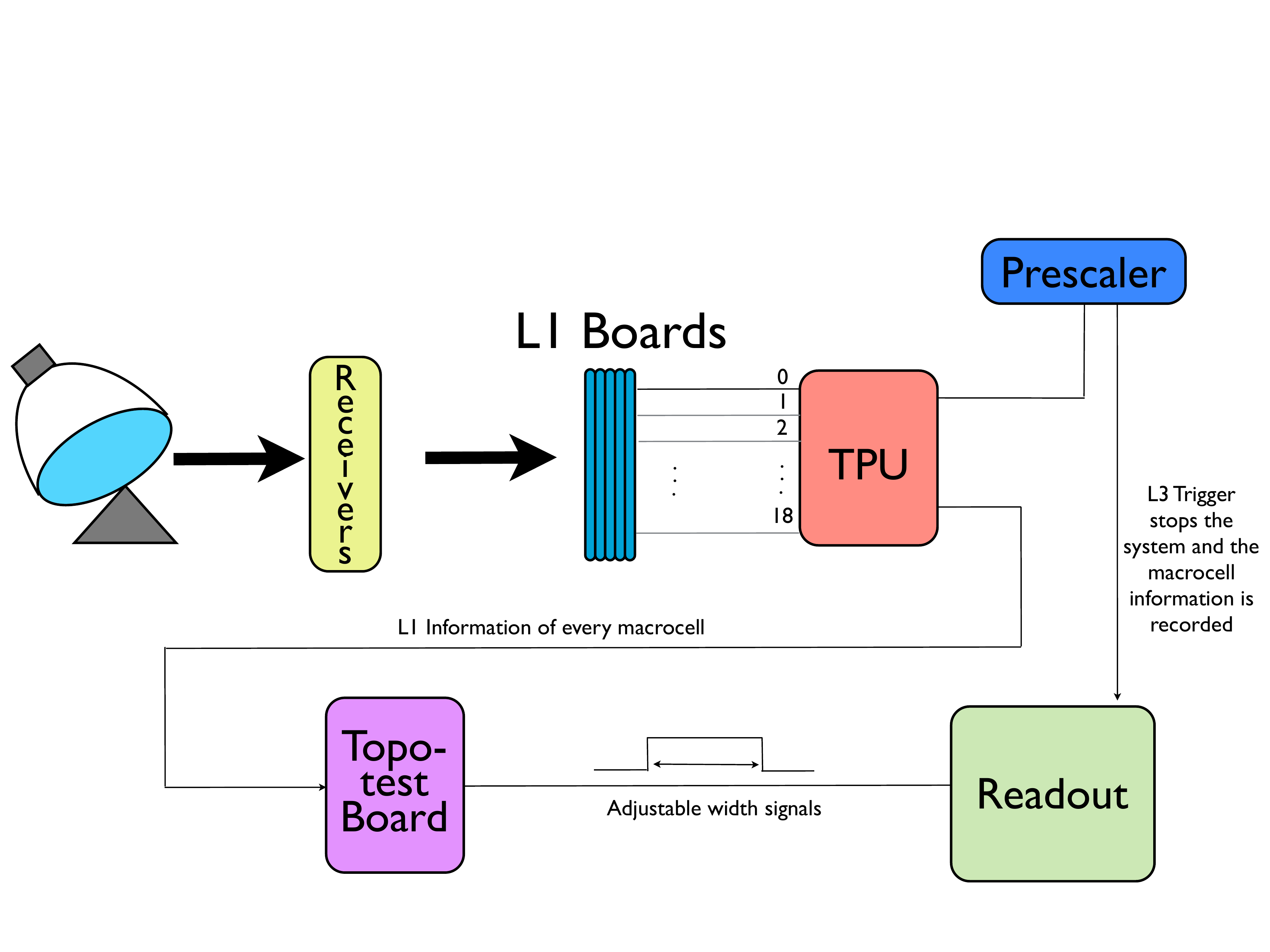}
\caption{Scheme of the recording system installed at the MAGIC telescopes.}
  \label{macrocell_recording}
 \end{figure}

We prepared a setup that records the digital output of the macrocells for every event recorded by the DAQ system \cite{DAQ_Diego}. A scheme of the setup used for recording the macrocell information is shown on Figure \ref{macrocell_recording}. 

\subsection{Sensitivity and $\gamma$-ray rate comparison}

We took Crab Nebula data at Az angles ranging from 100$^\circ$ to 175$^\circ$ and Zd between 6$^\circ$ and 22$^\circ$. We analyzed these data using the standard analysis in MAGIC and applied the standard $\gamma$/hadron separation and event reconstruction cuts for two energy ranges: medium to high energies and low energies.

\paragraph{Medium to high energies}

We applied the standard $\gamma$/hadron separation, reconstructed energy and $\theta^2$ cuts used for medium energies, the energy range where we achieve the best sensitivity. These cuts lead to an energy threshold of $\sim$ 250 GeV. We calculated the sensitivity of the telescope and the $\gamma$-ray rate for the sample without applying any cut in macrocells and applying the Topo macrocell cuts. We obtain that the ratio between the $\gamma$-ray rate obtained applying cuts and without applying them is:

$$\genfrac{}{}{0pt}{}{\gamma\mathrm{\text{-}ray\ rate}} {\mathrm{ratio}}=\frac{\gamma\mathrm{\text{-}ray\ rate} \mathrm{ [macrocell\ cuts]}}{\gamma\mathrm{\text{-}ray\ rate} \mathrm{ [no\ cuts]}}=1$$

For the sensitivity we obtain a similar result:

$$\genfrac{}{}{0pt}{}{\mathrm{Sensitivity}} {\mathrm{ratio}}=\frac{\mathrm{Sensitivity} \mathrm{ [macrocell\ cuts]}}{\mathrm{Sensitivity} \mathrm{ [no\ cuts]}}=1$$

Therefore we confirm that we have exactly the same number of background and $\gamma$-like events after applying the Topo-trigger macrocell selection cuts. The ratios given do not have errors because we have exactly the same number of events for both cases.

\paragraph{Low energies}

We also made an analysis applying the standard cuts for low energies (energy threshold $\sim$70 GeV).  We computed the ratio between the sensitivity with and without macrocell cuts and $\gamma$-ray rate with and without macrocell cuts as well. The result for the sensitivity ratio is:

$$\genfrac{}{}{0pt}{}{\mathrm{Sensitivity}} {\mathrm{ratio}}=1.005 \pm 0.007$$

And for the $\gamma$-ray rate we obtain a similar result:

$$\genfrac{}{}{0pt}{}{\gamma\mathrm{\text{-}ray\ rate}} {\mathrm{ratio}}=0.997 \pm 0.001$$

The results are summarized on Table \ref{Crab_summary}. These results confirm what we expected from the simulations: if we apply the macrocell selection to the events used for analysis, we basically keep the same detection efficiency. 

\begin{table}[h!tb]
\begin{center}
\begin{tabular}{c|c|c}
& Sensitivity ratio & $\gamma$-ray rate ratio\\
 \hline
 \hline
 Medium E   & 1  &  1  \\
 \hline
Low E  & 1.005 $\pm$ 0.007 & 0.997 $\pm$ 0.001 \\
\end{tabular}
\end{center}
\caption{Summary of the results for the sensitivity ratio and $\gamma$-ray rate ratio between the analysis applying the Topo-trigger macrocell cuts and without applying it for medium and low energies.}
\label{Crab_summary}
 \centering
\end {table}

\section{Discussion and conclusions}
\label{discussion}

We developed a novel stereo trigger system for IACTs which makes use of the topological information of the showers in the camera. Combining the information of the Az angle at which the telescope is pointing and the L1 trigger macrocell hit in each telescope we can reject 85\% of the accidental stereo trigger rate, which is the dominant at the lowest energies, without losing gamma rays. By studying the effect of applying the selection algorithm to off-axis data, we find that the discrimination power of the algorithm does not depend on the source position on the camera. We run simulations reducing the DT used for triggering the telescopes and applying this algorithm. We found that implementing this trigger translates into a decrease of up to 8\% in the energy threshold and an increment of  about 60\% in the collection area at the lowest energies and from 10--20\% at the energy threshold, where most of the events are triggered. The algorithm developed allows to lower DTs without increasing the accidental rate. We installed a device to record the triggered macrocells of the events recorded by the MAGIC telescope. Without reducing the DT applied at the L0 trigger, we verified that the Topo-trigger macrocell selection tested in the MC does not lead to any loss in sensitivity or in $\gamma$-ray rate. The potential of this easy to implement algorithm is that it could also be used for any other trigger based on macrocells or clusters where the position of the image in the camera is available \cite{LST_trigger}. 

The board that will be used to veto signals from the L3 trigger is already installed in the MAGIC telescope and is currently under commissioning.


\

\acknowledgments

We gratefully acknowledge the MAGIC collaboration for allowing us to perform tests at the telescopes site, and use their data analysis software. We would also like to thank the Instituto de Astrof\'{\i}sica de Canarias for the excellent working conditions at the Observatorio del Roque de los Muchachos in La Palma. Funding for this work was partially provided by the Spanish MINECO under projects FPA2012-39502, CPAN CSD2007-00042, and SEV-2012-0234. RL acknowledges support of the Otto-Hahn group at the Max Planck Institute for Physics.

\



\setcounter{table}{0}

\newpage
\appendix

\section{Topo-trigger selection tables depending on the Azimuth}
\label{super-topo tables}

\addtocounter{table}{+1}

\begin{table*}[h!tbp]
    \begin{minipage}{15.5cm}
\begin{center}
\begin{tabular}{c|c|c|c|c|c|c}
\hline
\hline
& \multicolumn{6}{|c}{ Macrocells selected in M2 } \\
 \hline
 
 Az cuts   &  19-49 & 49-79 & 79-109 & 109-139 & 139-169 & 169-199   \\
 \hline
M~I Macr.& \multicolumn{6}{|c}{  } \\
 \hline

   0   & 0, 1, 2  & 0, 1, 2, 6 & 0, 1, 6  & 0, 1, 5, 6 & 0, 5, 6 & 0, 5, 6  \\
 1    &  1, 7, 8 & 1, 7, 8, 18  & 1, 7, 18 & 1, 7, 18 & 1, 6, 18 &  1, 6, 18    \\
   2   & 2, 8, 9  & 1, 2, 8, 9 & 1, 2, 8  & 0, 1, 2, 8 & 0, 1, 2  & 0, 1, 2, 3      \\
   
   3      & 2, 3, 10  & 0, 2, 3, 10 & 0, 2, 3 &   0, 2, 3 &  0, 3, 4 &  0, 3, 4     \\
   4   & 0, 3, 4 & 0, 3, 4, 5 & 0, 4, 5  &  0, 4, 5 &  4, 5, 14 & 4, 5, 14    \\
    5  & 0, 5, 6  & 0, 5, 6 & 5, 6, 16 & 5, 6, 16 & 5, 15, 16 & 5, 15, 16    \\

  6    & 1, 6, 18  & 1, 6, 17, 18 & 6, 17, 18 & 6, 17, 18 & 6, 16, 17 &  5, 6, 16, 17  \\
 7    &  7, 8 & 7& 7 & 7, 18 & 7, 18 &  7, 18   \\
   8   & 8  &7, 8 & 7, 8 & 7, 8  & 1, 7, 8 & 1, 2, 7, 8  \\
    
    9     & 9  & 8, 9 & 8, 9 & 8, 9 &  2, 8, 9 & 2, 8, 9, 10   \\
   10   & 9, 10 & 9, 10 & 2, 9, 10 & 2, 9, 10 & 2, 3, 9, 10 & 2, 3, 10, 11  \\
     11 & 10, 11  & 10, 11 &  3, 10, 11 & 3, 10, 11, 12 &  3, 11, 12  & 3, 11, 12  \\
   
   12   & 3, 11, 12  & 3, 4, 11, 12 & 3, 4, 11, 12 & 3, 4, 12, 13 & 4, 12, 13 & 4, 12, 13   \\
 13    &  4, 12, 13 & 4, 12, 13, 14 & 4, 13, 14 & 4, 13, 14 & 13, 14 & 13, 14   \\
    14  & 4, 5, 13, 14  & 4, 5, 14, 15 & 5, 14, 15 & 5, 14, 15 & 5, 14, 15 & 14, 15    \\
   
    15     & 5, 15, 16 & 5, 15, 16 & 15, 16 & 15, 16 & 15, 16 & 15  \\
   16   & 6, 16, 17 & 6, 16, 17 &16, 17  & 16, 17 & 16, 17 & 15, 16   \\
     17 & 17, 18  & 17, 18 & 17, 18 &  17 & 17 & 16, 17   \\

    18  & 7, 18  & 7, 18 & 18 &17, 18 & 17, 18 & 17, 18     \\

\hline
\end{tabular}
\end{center}
\caption{Topo-trigger macrocells selected depending on the Azimuth. }
\label{Macrocells_selected}
    \end{minipage}
      \centering
\end {table*}

\addtocounter{table}{-1}


\begin{table*}[h!tbp]
    \begin{minipage}{15.5cm}
\begin{center}
\begin{tabular}{c|c|c|c|c|c|c}
\hline
\hline
& \multicolumn{6}{|c}{ Macrocells selected in M2 } \\
 \hline
 
 Az cuts    & 199-229 & 229-259 & 259-289 & 289-319 & 319-349 & 349-19  \\
 \hline
M~I Macr.& \multicolumn{6}{|c}{  } \\
 \hline

   0     & 0, 4, 5 & 0, 3, 4, 5 & 0, 3, 4 & 0, 2, 3, 4 & 0, 2, 3  & 0, 2, 3 \\
 1     &  0, 1, 6 & 0, 1, 2, 6   & 0, 1, 2 & 0, 1, 2 & 1, 2, 8 & 1, 2, 8   \\
   2   &  0, 2, 3 & 0, 2, 3 & 2, 3, 10 &  2, 3, 9, 10 & 2, 9, 10 & 2, 8, 9, 10    \\
   
   3     &  3, 4, 12 & 3, 4, 12  &  3, 11, 12 & 3, 11, 12 & 3, 10, 11 & 2, 3, 10, 11    \\
   4   & 4, 13, 14 &   4, 12, 13, 14 &   4, 12, 13 &  4, 12, 13  &   3, 4, 12 & 3, 4, 12  \\
    5   & 5, 14, 15 & 4, 5, 14, 15 & 4, 5, 14 & 0, 4, 5, 14  & 0, 4, 5, 14 & 0, 4, 5, 6   \\

  6    & 5, 6, 16 & 0, 5, 6, 16 & 0, 5, 6 &  0, 5, 6 &  0, 1, 6 & 0, 1, 6 \\
 7      &  1, 7, 18 &  1, 7, 8, 18 &  1, 7, 8 & 1, 7, 8 &  7, 8 &  7, 8  \\
   8    & 1, 2, 7, 8 & 1, 2, 8, 9 & 2, 8, 9 & 2, 8, 9 &8, 9 & 8, 9  \\
    
    9     & 2, 9, 10 & 2, 9, 10 & 9, 10  & 9, 10 & 9, 10 & 9 \\
   10     & 3, 10, 11 &  3, 10, 11  &10, 11  & 10, 11 & 10 & 9, 10  \\
     11  & 11, 12 & 11, 12 & 11, 12  & 11 &  11 & 10, 11   \\
   
   12      & 12, 13 & 12, 13 & 12 & 11, 12 & 11, 12 & 11, 12  \\
 13      & 13, 14 & 13 & 13 & 12, 13 & 12, 13 & 12, 13  \\
    14   & 14 & 13, 14 & 13, 14 & 13, 14 & 4, 13, 14 & 4, 13, 14  \\
   
    15      & 15 & 14, 15 & 14, 15 & 14, 15 & 5, 14, 15 & 5, 14, 15, 16  \\
   16     & 15, 16 &  15, 16 & 5, 15, 16 & 5, 15, 16 & 5, 6, 15, 16 & 5, 6, 16, 17 \\
   17   & 16, 17 & 6, 16, 17 & 6, 16, 17 & 6, 16, 17, 18 & 6, 16, 17, 18 & 6, 17, 18 \\

    18   & 6, 17, 18 & 1, 6, 17, 18 & 1, 6, 17, 18 & 1, 6, 7, 18 & 1, 7, 18 & 1, 7, 18   \\
\hline
\end{tabular}
\end{center}
\caption{Continued}
\label{Macrocells_selected2}
    \end{minipage}
      \centering
\end {table*}

\end{document}